\documentclass[aps,prb,groupaddress,twocolumn,floatfix]{revtex4-2}
\usepackage{units}
\usepackage{amsmath}
\usepackage{amssymb}
\usepackage{graphicx}
\usepackage{bm}
\usepackage{multirow,xcolor,relsize,ulem,microtype}

\newcommand{\be}{\begin{equation}}
\newcommand{\ee}{\end{equation}}
\newcommand{\re}[1]{\text{Re}[#1]}
\newcommand{\im}[1]{\text{Im}[#1]}

\begin{document}

\title{Observing complementary Lucas sequences using non-Hermitian zero modes}

\author{Li Ge}
\email{li.ge@csi.cuny.edu}
\affiliation{\textls[-18]{Department of Physics and Astronomy, College of Staten Island, CUNY, Staten Island, New York 10314, USA}}
\affiliation{The Graduate Center, CUNY, New York, New York 10016, USA}

\date{\today}

\begin{abstract}
The Lucas sequences are integers defined by a homogeneous recurrence relation. They include the well-known Fibonacci numbers, which appear abundantly in nature. The complementary Lucas numbers, defined by the same recurrence relation, are less well-known. In this work, we show that such complementary Lucas sequences can be observed simultaneously on the same physical platform, using zero modes in one-dimensional system(s) coupled to a non-Hermitian reservoir. When the latter consists solely of gain and loss modulated non-Hermiticity, the two Lucas sequences can manifest as a linearly localized edge state and a constant-intensity mode, respectively. When the reservoir features asymmetric couplings as well, more general Lucas sequences can be realized either directly on one sublattice in the reservoir or by reconstructing the wave function on both sublattices. Finally, if we allow, in addition, an alternate frequency detuning in the reservoir, \textit{all} Lucas sequences can be observed in principle, including the Fibonacci and Lucas numbers.
\vspace{30pt}

\end{abstract}

\maketitle

\section{Introduction}
\vspace{-10pt}
Recurrence relations have found a wide range of applications in physics, chemistry, biology, computer science, and economics \cite{MathBook}. For example, the transfer matrix method \cite{Kramer,Yeh} is a simple yet powerful technique for calculating wave propagation in physical systems. It uses a linear map to relate the values of the wave function on consecutive planes. The Fibonacci numbers, satisfying $f_n=f_{n-1}+f_{n-2}\,(n\in \mathbb{Z})$, present another familiar example of such recurrence relations, which have been used in physics to construct, for example, quasi-periodic potentials \cite{Kohmoto} that feature unique spectral characteristics and scaling behaviors \cite{Macia,Jagannathan,Bonsel}.

In a recent study \cite{LL}, a linear recurrence relation was shown to be essential for a ``linear localization" phenomenon in coupled photonic systems \cite{NPreview}, which was exhibited by a symmetry and topologically protected zero mode \cite{Hasan,Qi,Alicea,Beenakker}: When weakly coupled to a one-dimensional non-Hermitian lattice with gain and loss modulation (``reservoir''), an edge state with zero energy (``zero mode'') in a Hermitian lattice (``system'') can display a linearly decreasing amplitude as a function of space in the reservoir. This linear waveform occurs when the zero mode transitions from the exponentially localized regime \cite{Localization_RMP1,Localization_RMP2,Localization_PT} to the delocalized regime in the reservoir; %and it is not the consequence of an exceptional point (EP) in a Hamiltonian, either of the coupled lattice or the reservoir itself. Instead, it is due to the zeroness of the eigenstate energy (i.e., $E\approx0$), partially warranted by the non-Hermitian particle-hole (NHPH) symmetry of the coupled lattice \cite{zeromodeLaser,NHFlatband_PRL,NHFlatband_PRJ,kawabata,Okugawa,Wu}, and 
at which point the recurrence relation of its wave function on both even-numbered and odd-numbered lattice sites in the reservoir (``sublattices''), i.e.,
\be
\Psi_n = 2\alpha \Psi_{n-2} - \Psi_{n-4}\quad (\alpha\in\mathbb{R}),\label{eq:iter}
\ee
becomes simply 
\be
\Psi_n = 2 \Psi_{n-2} - \Psi_{n-4}.\label{eq:iter2}
\ee

Equation (\ref{eq:iter2}) is a special example of the linear recurrence relation
\be
F_m = PF_{m-1}-QF_{m-2}\quad (P,Q\in\mathbb{Z}) \label{eq:lucas}
\ee
that defines the Lucas sequences. For a given pair of $P$ and $Q$, Eq.~(\ref{eq:lucas}) allows two linearly independent and complementary integer solutions. They can be differentiated by their two initial numbers: One starts with 0 and 1 by convention, and the other 2 and $P$. The most famous example is the Fibonacci numbers 0, 1, 1, 2, 3, 5, $\ldots$ with $P=1$ and $Q=-1$. Its complementary solution with the same $P$ and $Q$, on the other hand, is the lesser-known Lucas numbers 2, 1, 3, 4, 7, 11, $\ldots$

In this article, we first show that the two complementary Lucas sequences of Eq.~(\ref{eq:iter2}) can be observed, up to an overall phase and amplitude choice, on the same physical platform. They are manifested as linearly localized edge states already mentioned and a constant-intensity mode \cite{CI1,CI2,activeResonance}, respectively. 

For the former, the previous finding is limited to the weak coupling limit of the system and the reservoir, leading to a weak linear tail in the reservoir compared to the peak amplitude of the edge state in the system \cite{LL}. 
Here, we first lift this severe limitation by properly engineering the non-Hermitian landscape of the coupled lattice: By allowing the system to be also non-Hermitian, i.e., including loss on the system side as well, we can achieve linear localization in the strong system-reservoir coupling regime, which enhances the linear tail in the reservoir and facilitates its potential experimental observation. We further study a different configuration where the non-Hermitian reservoir bridges two systems%(``system 1'' and   ``system 2'')
, and we show that linear localization can be observed, for example, when the second system supports a zero mode with a vanished amplitude at the coupling site.

While energy eigenstates displaying a constant intensity may seem common in a ring geometry due to its rotation symmetry and traveling-wave solutions, the energy degeneracy between a pair of clockwise and counterclockwise modes often leads to the observation of standing-wave patterns instead \cite{Zhang2018}. Approaches to lift this energy degeneracy include asymmetric scattering or coupling induced exceptional points \cite{Feng2014,Wiersig,Peng,Hayenga,Zhang2022}, nonlinear bifurcations \cite{Cao}, and spin-orbital couplings \cite{Sala}. A constant intensity is even more elusive in a linear and Fabry-Perot-like structure. Previously identified systems with this property require not only a modulation of the imaginary part of the on-site potential (or refractive index in optical setups) but also its real part \cite{CI1,CI2}. An exception is found in an open system with outgoing boundary conditions \cite{activeResonance}, where only modulating the real part of the refractive index is required. Here, we show that the complementary solution of linear localization in the recurrence relation (\ref{eq:iter2}) provides another exception, where only a modulation of the \text{imaginary} part of the on-site potential is required. Furthermore, we show that a constant phase on each sublattice coexists with the constant intensity.

Last but not least, we show that the general recurrence relation (\ref{eq:lucas}) can be realized on each sublattice of our non-Hermitian reservoir, by including asymmetric couplings and frequency detunings on top of alternate gain and loss \cite{NHchiral}. With an open boundary condition on the right end of the reservoir, we identify a single Lucas sequence on one sublattice (such as the Mersenne numbers with $P=3,Q=2$ and the Fibonacci numbers with $P=1=-Q$) and the simultaneous excitation of the two complementary Lucas sequences in each case on the other sublattice.  

\section{Lucas sequences}
\vspace{-10pt}

The Lucas sequences $F_{m}$ defined above by Eq.~(\ref{eq:lucas}) have two explicit solutions given by  
\be
U_m %= \frac{a^m-b^m}{a-b} 
= \sum_{n=0}^{m-1}a^nb^{m-n-1}, \quad V_m = a^m+b^m, \label{eq:Vm}
\ee
where $m$ is a non-negative integer, $a,b=(P\pm\sqrt{D})/2$, and $D = P^2-4Q$. It is easy to check that these solutions indeed start with $0$ and $1$ (for $U_m$) and with $2$ and $P$ (for $V_m$), as we mentioned in the introduction. With $P=1$ and $Q=-1$, we have $a,b=(1\pm\sqrt{5})/2$, and $U_m,V_m$ give the Fibonacci numbers and Lucas numbers, respectively. 

When $a,b$ become the same and equal $s\neq0$, one finds
\be 
U_m(P,Q) = m s^{m-1},\quad V_m(P,Q) = 2s^m. \label{eq:repeated}
\ee
If we further let $s=1$, or equivalently, $P=2,Q=1$ in Eq.~(\ref{eq:lucas}) and $\alpha=1$ in Eq.~(\ref{eq:iter}), these two complementary sequences become
\be 
U_m(P,Q) = m,\quad V_m(P,Q) = 2, \label{eq:repeated2}
\ee
from which linear localization and a constant-intensity mode seem bound to arise, respectively. 

This is, however, not the case. %We first note that despite the periodic construction of our reservoir (i.e., with alternate gain and loss), the non-Bloch solution \cite{Bloch} given by Eq.~(\ref{eq:repeated2}) are possible due to a single root of multiplicity 2 in the characteristic equation of Eq.~(\ref{eq:iter}). 
We note that the step size of the two sequences given by Eq.~(\ref{eq:iter}) is 2 in terms of the index $m$ in Eq.~(\ref{eq:lucas}). As a result, the wave function forms two separate linear relations in the reservoir in general, one on the even-numbered sublattice and the other on the odd-numbered sublattice. Each of these linear relations can be a different superposition of $U_m(P,Q)$ and $V_m(P,Q)$ in Eq.~(\ref{eq:repeated2}), meaning that they have different slopes with respect to the lattice position on these two sublattices \textit{and} different vertical shifts. In other words, these two linear relations are not necessarily aligned to cause either linear localization or a constant-intensity mode.

\section{Linear localization}

To realize linear localization in a gain and loss modulated reservoir, it was found that an additional requirement must be satisfied, i.e., the energy of this mode in the coupled system-reservoir lattice needs to be $E=0$ \cite{LL}. 
% The latter leads to $\gamma=2t$ in the reservoir, which in turn leads to the boundary condition $\Psi_{N-1} = 2i\Psi_N$ and $2i\Psi_{N-1} + \Psi_N + \Psi_{N-2}=0$. 

This condition can be warranted by either chiral symmetry \cite{NHchiral,kawabata,Okugawa,Wu} or pseudo-chirality \cite{pseudoChirality} for a symmetry-protected zero mode in a non-Hermitian system. With either symmetry, the complex spectrum of the entire lattice is symmetric about the origin of the complex energy plane, i.e., satisfying $E_\mu = -E_\nu$. A zero mode with $E=0$ emerges when the two mode indices $\mu,\nu$ are identical. 

\begin{figure}[t]
  \centering \includegraphics[width = \linewidth]{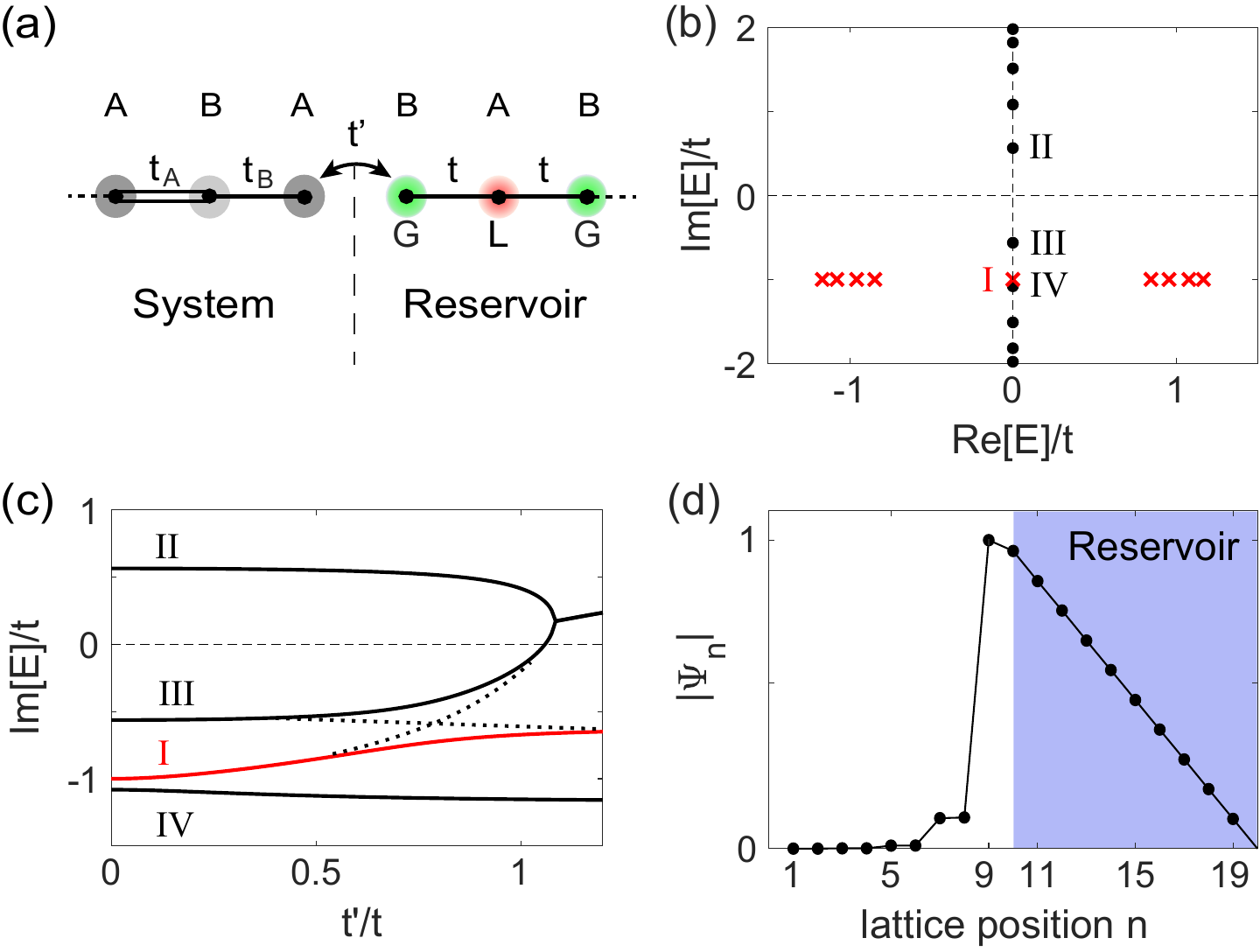}
  \caption{(a) Schematic showing the coupling between the system and the reservoir. (b) Spectra of the system (red crosses; 9 sites) and the reservoir (black dots; 10 sites) before the coupling. $\gamma=2t$ is used in the reservoir and $\kappa_0=t_B=t$, $t_A=0.2t$ in the system. (c) $\im{E}$ of the four modes marked by I to IV in (b). Dashed line shows $\im{E}=0$ and dotted lines indicate the avoided crossing between modes I and III. (d) Wave function of mode III with $E=0$ at $t'=1.06t$ (dots). Shaded area indicates the reservoir. Solid line guides the eye and shows the linear tail in the reservoir. 
  }\label{fig:schematic}
\end{figure}

These two symmetries, however, are strongly restrictive in non-Hermitian systems. For example, one approach to realize non-Hermitian chiral symmetry is by using the Clifford algebra, which is difficult to identify and construct beyond systems built upon Pauli or Dirac matrices \cite{NHchiral}. Another approach to realize non-Hermitian chiral symmetry is by combining non-Hermitian particle-hole (NHPH) symmetry \cite{zeromodeLaser,NHFlatband_PRL,NHFlatband_PRJ} and parity-time (PT) symmetry \cite{NPreview}. They are defined by the anti-commutation relation $\{H,CT\}=0$ and commutation relation $[H,PT]=0$ respectively, where $T$ is time reversal in the form of a complex conjugation and $C,P$ are two linear operators. They warrant a spectrum symmetric about the imaginary and real axes, respectively. Together, they then lead to a spectrum symmetric about the origin $E=0$, accompanied by the chiral symmetry $\{H,CP\}=0$. This approach, while more universal, is difficult to materialize with two coupled lattices, where the distinct system and reservoir [see Fig.~\ref{fig:schematic}(a)] eliminate a potential PT symmetry even when $P$ is not restricted to parity.

Nevertheless, this observation reveals a different strategy to realize a zero mode with $E=0$: One first uses a symmetry to warrant $\re{E}=0$ and then tunes the system to have $\im{E}=0$ as well. The former can be warranted by NHPH symmetry alone, which leads to $E_\mu = -E_\nu^*$. A non-Hermitian zero mode, defined by $\re{E_\mu}=0$, emerges when $\mu=\nu$. Unlike chiral symmetry or pseudo-chirality, NHPH symmetry is a general property of many gain and loss modulated lattices \cite{zeromodeLaser}, including the coupled lattice that exhibits linear localization. To further satisfy $\im{E}=0$ (and $E=0$) \textit{approximately}, Ref.~\cite{LL} opted for a weak coupling $t'$ between the system and the reservoir, which slightly perturbs the energy of the Hermitian zero mode in the system along the imaginary axis. 

This requirement of a weak system-reservoir coupling $t'$ imposes a severe limitation on the linear tail of the non-Hermitian zero mode: It only features a weak linear tail in the reservoir, with its peak amplitude about one order of magnitude lower than that in the system. Increasing the coupling $t'$ enhances the weight of the non-Hermitian zero mode in the reservoir, but its wave function no longer displays a linear tail, as the requirement $\im{E}\approx0$ breaks down. 

\subsection*{Strong coupling regime}
To overcome this issue in the strong coupling regime, we first observe that with the system coupled to a gain site in the reservoir [see Fig.~\ref{fig:schematic}(a)], linear localization leads to net gain if the system is kept Hermitian, simply because the maximum amplitude of the linear tail is at a gain site in the reservoir (i.e., the one the system couples to). 

To offset this net gain, one may attempt to apply an additional uniform loss across both the system and reservoir. This approach, however, does not change the wave functions of the coupled lattice, because it is equivalent to shifting the origin $E=0$ upward in the complex plane. In terms of Eq.~(\ref{eq:iter}), the condition $\alpha=1$ actually requires that $\im{E}$ equals the average of the uniform gain in one sublattice of the reservoir and the uniform loss in the other sublattice:
\be 
\im{E} = \frac{\im{V_B} + \im{V_A}}{2}.\label{eq:ave}
\ee
$V_{A,B}$ are the on-site potentials of the two sublattices in the reservoir, with $\re{V_{A,B}}=0$ as in the system. 

In the original configuration with a Hermitian system, the gain (loss) is given by $\im{V_B}=\gamma$ ($\im{V_A}=-\gamma$) in the reservoir, leading to the requirement $\im{E}=0$. Now with an additional loss (denoted by $\kappa_0$), the left hand side of Eq.~(\ref{eq:ave}) can be reduced to 0 in the strong coupling regime, but the right hand side of Eq.~(\ref{eq:ave}) is also reduced by the same amount to $-\kappa_0$. As a result, Eq.~(\ref{eq:ave}) cannot be satisfied by imposing additional uniform loss across both the system and the reservoir in the strong coupling regime.

However, Eq.~(\ref{eq:ave}) can be satisfied in the strong coupling regime by introducing additional loss \textit{only in the system}. As we exemplify in Fig.~\ref{fig:schematic}, an imaginary on-site potential $V_0=-i\kappa_0$ is added to an otherwise Hermitian Su-Schrieffer-Heeger (SSH) chain \cite{SSH,SSH2}, which is used as our system:
\be
H_S  = -i\sum_n \kappa_0 |n\rangle \langle n|  + \left(\,t_n|n+1\rangle \langle n|  + h.c.\,\right). \label{eq:Hs}
\ee
Here the nearest-neighbor (NN) coupling $t_n$ alternate between $t_A$ (when $n$ is odd) and $t_B$ (when $n$ is even). With $t_A>t_B>0$, a single topologically protected zero mode is localized at the right edge of the system when there are an odd number of lattice sites in the system (e.g., nine in Fig.~\ref{fig:schematic}). 

Next, we couple this system (and its zero mode) by NN coupling $t'$ to a non-Hermitian reservoir on the right, described by 
\be
H_R  = \sum_n V_n |n\rangle \langle n|  + \left(\,t|n+1\rangle \langle n|  + h.c.\,\right). \label{eq:Hr}
\ee
$t>0$ is its uniform NN coupling, and $V_n$ is the imaginary on-site potential representing gain and loss as mentioned previously. It is given by $V_B=i\gamma$ when $n$ is even and $V_A=-i\gamma$ when $n$ is odd. We note here that $n$ is counted continuously from the left of the system to the right of the reservoir. $\alpha$ in Eq.~(\ref{eq:iter}) is related to $\gamma$ by \cite{LL}
\be
\alpha = \frac{\gamma^2}{2t^2}-1 \label{eq:gamma}
\ee
in a zero mode with $E=0$. Therefore, $\alpha=1$ for linear localization is satisfied by $\gamma=2t$. 

Before we introduce the coupling $t'$, the system and reservoir have two distinct spectra. At the linear localization condition $\gamma=2t$ mentioned above, the spectrum of the reservoir is on the imaginary axis,  while that of the system is on the horizontal line given by $\im{E}=-\kappa_0$ [see Fig.~\ref{fig:schematic}(b)]. When $\kappa_0$ is sufficiently large, i.e., comparable to the coupling $t$ in the reservoir, there is no mode in the vicinity of $E=0$, including the topological edge mode of the SSH chain (mode I). 

As we increase $t'$, one mode with $E=0$ emerges, as we show in Fig.~\ref{fig:schematic}(c). This zero mode is given by mode III at $t'/t\approx1.06$, and this strong coupling leads to a strong linear tail in the reservoir as can be seen clearly in Fig.~\ref{fig:schematic}(d). Note that this mode undergoes an avoided crossing with the original topological edge mode I near $t'/t=0.8$, and hence it inherits the latter's topological property and exponential-like tail on the system side. In this process, all non-Hermitian zero modes close to $E=0$ (i.e., modes I to IV) remain on the imaginary axis, until modes II and III acquire a finite real part after coalescing at an exceptional point when $t'/t\approx1.09$ [see Fig.~\ref{fig:schematic}(c)].

\subsection*{A reservoir bridging two systems} 

\begin{figure}[t]
  \centering \includegraphics[width = \linewidth]{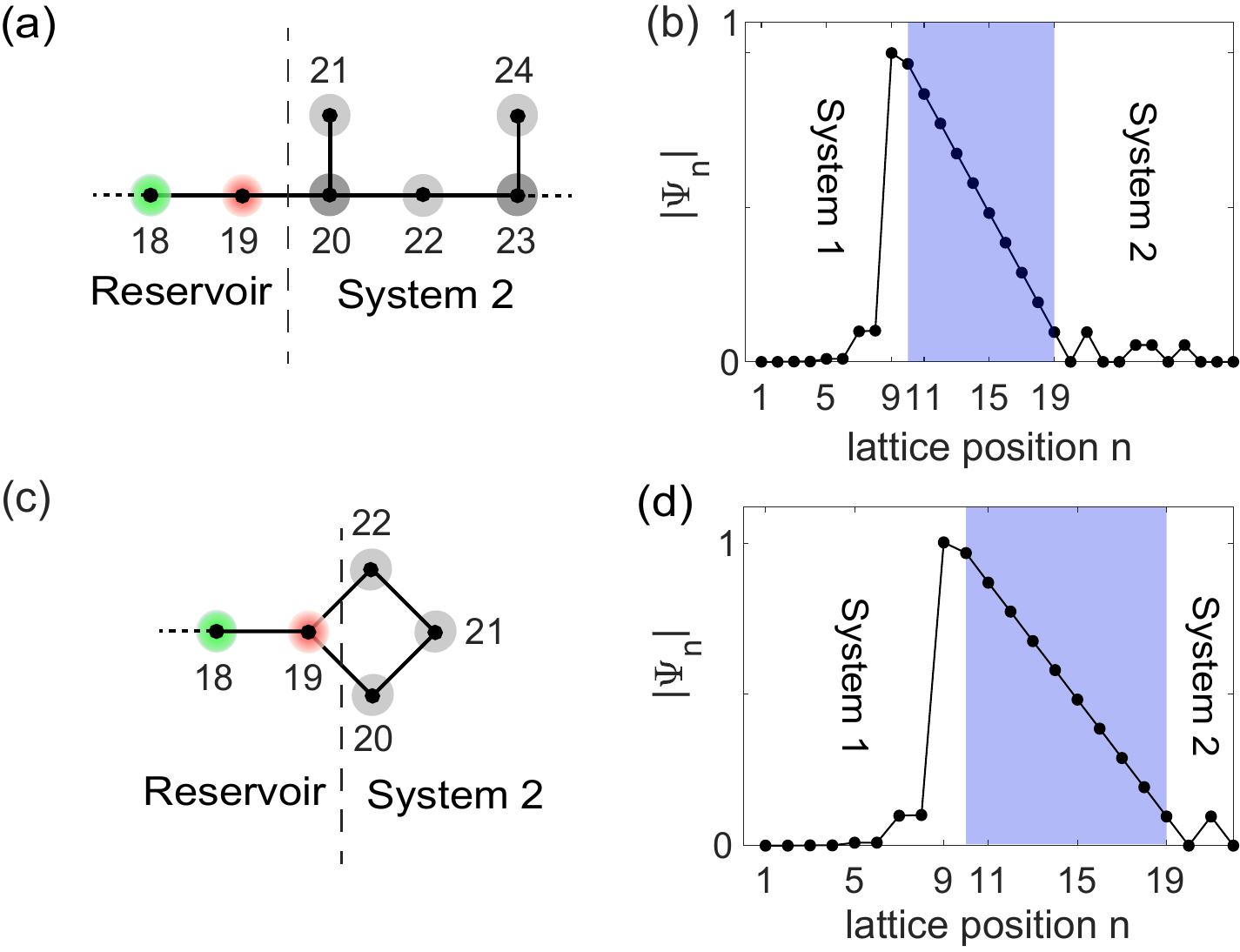}
  \caption{(a,c) Schematics showing the reservoir coupling to a second system on the right. The coupling(s) between them and inside system 2 are taken to be $t$ for simplicity. The Lieb lattice in (a) has 11 sites and terminates with a pair of vertically coupled sites on the right. Other parameters are the same as in Fig.~\ref{fig:schematic}. (b,d) Linear localization manifested by a corresponding zero mode in (a,c).
  }\label{fig:termination}
\end{figure}

In our discussions so far, the reservoir terminates with an open boundary condition on the right, i.e., the last site (number 19 in Fig.~\ref{fig:schematic}) is only coupled to its neighbor on the left. This boundary condition is most common in coupled photonic and acoustic platforms \cite{NPreview}, and mathematically, it is equivalent to adding one additional site on the right and imposing the Dirichlet boundary condition at this site \cite{Ge_PTlength}, e.g., $\Psi_{20}=0$ in Fig.~\ref{fig:schematic}. 

This effective boundary condition can also be achieved using canceled couplings when the right side of the reservoir couples to a second system (``system 2''), similar to the formation of a ``dark state'' in atomic physics. We show two such examples in Fig.~\ref{fig:termination}. 

In the first example, system 2 is a Lieb lattice \cite{Lieb,Lieb2} with 11 site [see Fig.~\ref{fig:termination}(a)]. It has three zero modes of its own, each formed by canceled couplings and with vanished amplitude at site 20. Therefore, the original zero mode shown in Fig.~\ref{fig:schematic}(d) is still a zero mode of the entire chain (with $\Psi=0$ everywhere in system 2), unaffected by the Lieb lattice. Realistically though, it is mixed with the three zero modes of the Lieb lattice due to their energy degeneracy, forming four zero modes in total, \textit{all} with a linear tail in the reservoir. One example is shown in Fig.~\ref{fig:termination}(b), where the vanished wave function at site 20 is due to the opposite values of the wave function at site 19 in the reservoir and site 21 in system 2.  

In the second example shown in Fig.~\ref{fig:termination}(c), system 2 also has a zero mode of its own, with $\Psi_{20} = -\Psi_{22}=1$ and $\Psi_{21}=0$. This zero mode, however, is destroyed when coupled to the reservoir. Nevertheless, the original zero mode shown in Fig.~\ref{fig:schematic}(d) now extends into system 2 [see Fig.~\ref{fig:termination}(d)], with $E=0$ and $\Psi_{20}=0$ still satisfied. The latter, as well as $\Psi_{22}=0$, is again the result of the canceled couplings from site 19 in the reservoir and site 21 in system 2.  

\section{Constant-intensity mode} 

\begin{figure}[t]
  \centering \includegraphics[width = \linewidth]{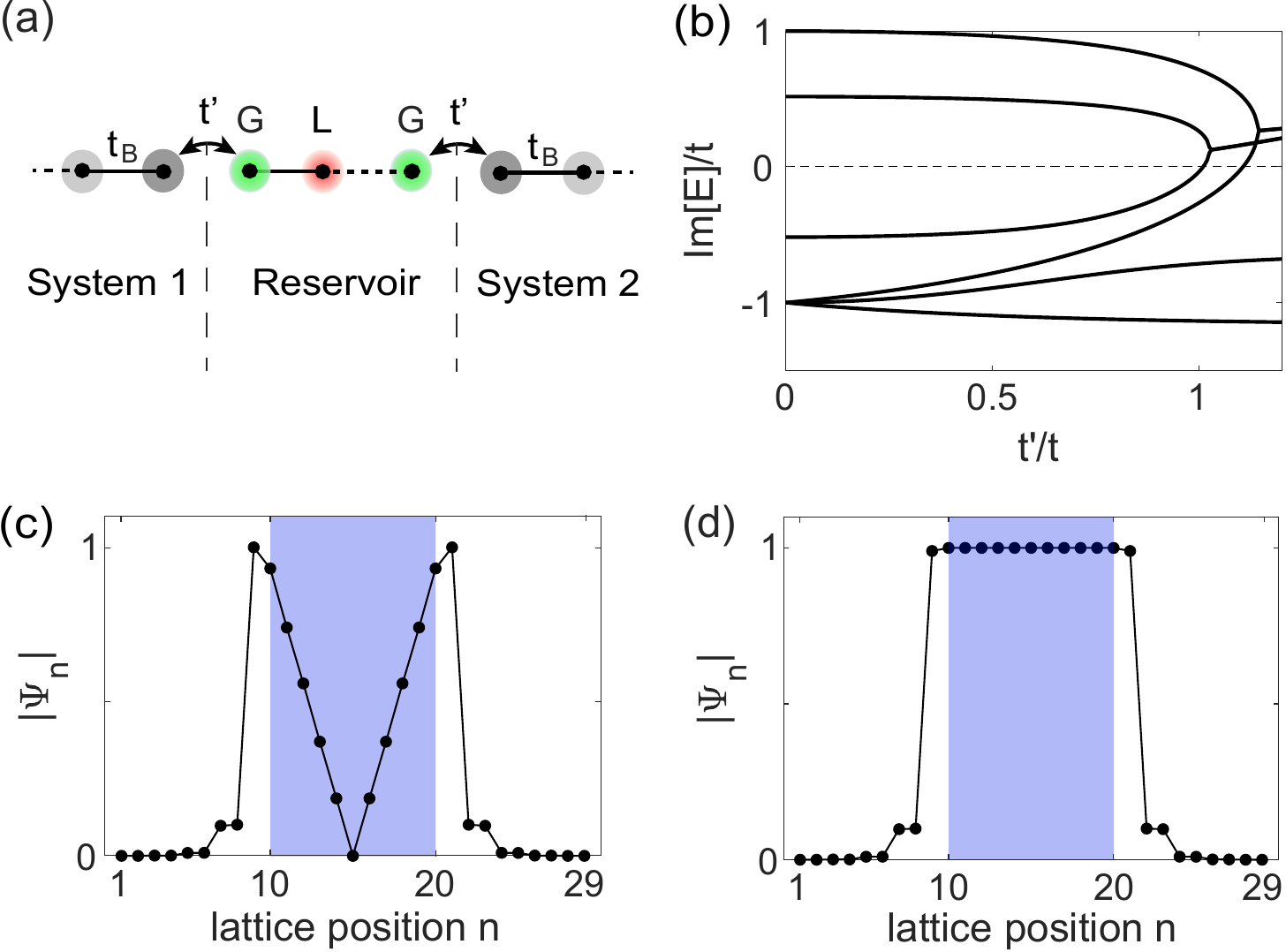}
  \caption{(a) Schematic showing two mirror-symmetric systems 1 and 2 bridged by the reservoir with an odd number of sites. (c) Evolution of $\im{E}$ with $t'$, showing two zero modes with $E=0$ at $t'/t\approx$ 1.01 and 1.11. Their spatial profiles are shown in (d) and (c), respectively. Other parameters are the same as in Fig.~\ref{fig:schematic} except that there are now 11 sites in the reservoir. }\label{fig:2systems}
\end{figure}

As we showed at the end of the previous section, the open boundary condition or its equivalence is critical for the realization of linear localization. This observation is also important for us to realize the complementary Lucas sequence, i.e., a constant-intensity mode based on $V_m$ in Eq.~(\ref{eq:repeated2}).

For this purpose, we remind the reader that the two complementary Lucas sequences given by Eq.~(\ref{eq:Vm}) have the same recurrence relation [i.e., Eq.~(\ref{eq:lucas})] and that they are determined by the first two numbers in the sequence. For example, they are 0, 1 in the Fibonacci numbers and 2, 1 in the Lucas numbers.

On our lattice, this difference is mapped to the boundary condition between the reservoir and the system(s). The constant solution $V_m$ in Eq.~(\ref{eq:repeated2}) indicates that a Neumann boundary condition (i.e., with vanished spatial derivative of the wave function) is required. This boundary condition, however, is difficult to realize on coupled photonic and acoustic platforms \cite{NPreview}. 

Nevertheless, we note that two such constant solutions exist side-by-side in general, one on each sublattice of the reservoir. To realize a constant-intensity mode, we want them to have the same amplitude but not necessarily the same phase. This property is indeed possible in a non-Hermitian zero mode warranted by the NHPH symmetry \cite{zeromodeLaser,NHFlatband_PRL,NHFlatband_PRJ}, where the phase difference between any two neighbors is $\pm\pi/2$. In other words, if such a zero mode displays a constant-intensity profile, we expect to find $\psi_n = \psi_{n-2} = \ldots \equiv C$ and $\psi_{n-1} = \psi_{n-3} = \ldots \equiv \pm iC$. Using the tight-binding model $H_R$ given by Eq.~(\ref{eq:Hr}), we derive
\be
-\frac{V_n}{t}\Psi_{n-1} = \Psi_{n} + \Psi_{n-2} = 2\Psi_{n}.
\ee
Here $V_n=\pm i\gamma = \pm 2it$ is required [see the discussion below Eq.~(\ref{eq:gamma})], and hence we look for an effective and phase-modified Neumann boundary condition $\Psi_n = \pm i\Psi_{n-1}$ at both ends of the reservoir.

One configuration where this effective boundary condition is guaranteed is a parity-symmetric lattice, i.e., with mirror-symmetric systems 1 and 2 bridged by our reservoir with an odd number of sites [see Fig.~\ref{fig:2systems}(a)]. The parity symmetry warrants that all modes of the lattice are either symmetric or anti-symmetric. When the conditions $\alpha=1$ and $E=0$ are satisfied in an anti-symmetric mode, i.e., with $\Psi=0$ at the center of the entire chain, we must have linear localization on each half of the reservoir [see Fig.~\ref{fig:2systems}(c)]. Similarly, when the same conditions are satisfied in a symmetric mode with no node in the reservoir, we must have a constant-intensity mode instead [see Fig.~\ref{fig:2systems}(d)].    

These conjectures are indeed confirmed, as we have seen in Figs.~\ref{fig:2systems}(c) and \ref{fig:2systems}(d), where the system-reservoir coupling $t'$ equals approximately 1.11$t$ and 1.01$t$, respectively [see Figs.~\ref{fig:2systems}(b)]. These two zero modes are the only ones found when we vary $t'$, with the other requirement $\alpha=1$ satisfied by fixing $\gamma$ at $2t$ [see Eq.~(\ref{eq:gamma})].   

\begin{figure}[b]
  \centering \includegraphics[width = \linewidth]{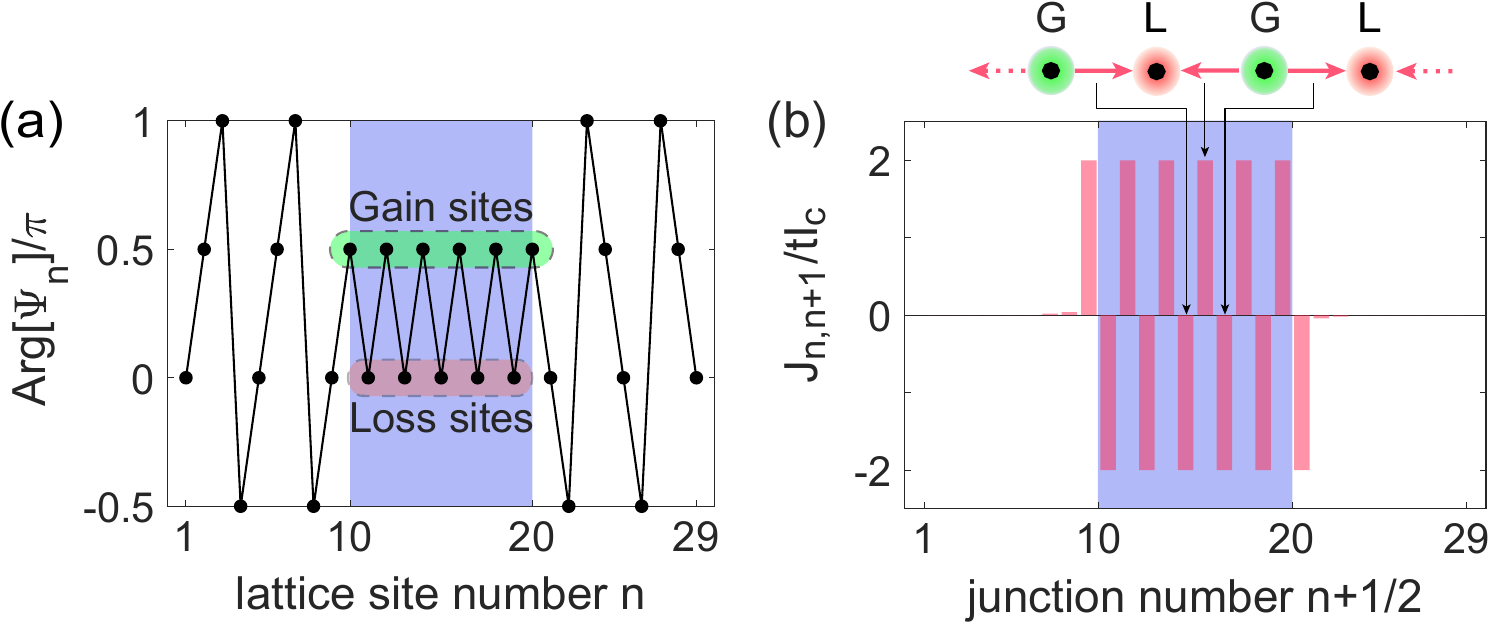}
  \caption{(a,b) Phase and energy flux of the constant-intensity mode shown in Fig.~\ref{fig:2systems}(d). In (a) the gain and loss sites in the reservoir are marked. In (b) the energy flux $J_{n,n+1}$ is plotted at the junction between sites $n$ and $n+1$, i.e., at $n+1/2$, and a negative current means the power flows from site $n$ to $n+1$ [see the inset].}\label{fig:CI}
\end{figure}

In addition to the constant intensity, $V_m$ in Eq.~(\ref{eq:repeated2}) also indicates a constant phase on each sublattice of the reservoir [see Fig.~\ref{fig:CI}(a)]. The phase of this zero mode in the two systems, on the other hand, winds linearly with the lattice position. This property is distinct from the underlying edge states of the two Hermitian SSH chains in systems 1 and 2, where the phase of these zero modes is undefined on one sublattice where their amplitude vanishes. Physically, this winding phase, similar to a traveling wave, indicates that energy flows continuously from the reservoir to the outer edges of the two systems [see our discussion of Fig.~\ref{fig:CI}(b) below]. 

The unique phase profile in the reservoir provides a physical understanding of the required loss in the two systems. Due to the zero mode property $E=0$, each lattice site has balanced energy input and output, consisting of on-site energy gain or loss and energy flux from one site to another. More specifically, the lattice Hamiltonian leads to
\be
\frac{d|\Psi_n|^2}{dt} = 2\im{V_n} |\Psi_n|^2 + J_{n,n+1} + J_{n,n-1}. \label{eq:continuity}
\ee 
The first term on the right is the on-site gain (loss) when $\im{V_n}$ is positive (negative). The other two terms are the energy fluxes from sites $n\pm1$ to site $n$ \cite{flux,Ge_PRB2023a}, defined by 
\be
J_{n,n\pm1}=it\Psi_{n\pm1}^*\Psi_n + c.c. \label{eq:flux}
\ee
These energy fluxes are defined similarly to the probability current in quantum mechanics. Besides the wave-energy flow, whether they also indicate mass transfers depends on the particle or quasi-particle described by the wave function.

Below we first consider an internal site $n$ in the reservoir away from the two systems. Sites $n\pm1$ on the other sublattice are both in the reservoir, and we find  $\Psi_{n+1}=\Psi_{n-1}$ (with the same phase and amplitude)  
in a constant-intensity mode, thanks to $V_m$ in Eq.~(\ref{eq:repeated2}). 
Therefore, we find $J_{n,n+1}=J_{n,n-1}$ as well.

In an energy eigenstate, the left hand side of Eq.~(\ref{eq:continuity}) equals $2\im{E} |\Psi_n|^2$, which vanishes in a zero mode because $E=0$. If $n$ is a gain site, then $J_{n,n\pm1}$ must be negative to cancel the on-site gain. In other words, this site provides gain in the form of energy fluxes to its two neighbors with loss. More quantitatively, we find
\be 
J_{n,n\pm1} = -\gamma|\Psi_n|^2\equiv -\gamma I_c\label{eq:balance}
\ee
from this energy balance, with $J_{n,n\pm1}$ each equal to half the on-site gain. $I_c$ here is the constant intensity in the reservoir. 

From the negative sign of $J_{n,n\pm1}$, we know that the phase of $\Psi_n$ at a gain site is $\pi/2$ ahead (instead of behind) of $\Psi_{n\pm1}$ on the loss sublattice [see Fig.~\ref{fig:CI}(a)]. Using this result, we then derive 
\be 
J_{n,n\pm1}=-2t|\Psi_{n}\Psi_{n\pm1}|=-2tI_c
\ee 
directly from Eq.~(\ref{eq:flux}). By comparing it with Eq.~(\ref{eq:balance}), we recover the condition $\gamma=2t$ (and $\alpha=1$) required by the recurrence relation (\ref{eq:iter2}). 

Next, we focus on a gain site at the edge of the reservoir, e.g., site 10 in Fig.~\ref{fig:CI}(a). The energy flux from its neighboring loss site in the reservoir (i.e., site 11) is negative and given by Eq.~(\ref{eq:balance}), but it is insufficient to balance out the on-site gain at site 10, which is twice as strong. The excessive energy must flow to system 1 and be balanced out there. Therefore, system 1 must have loss and so does system 2, due to the parity symmetry of the entire lattice. 

The equal division of energy flux from this edge gain site to its two neighbors, one in the reservoir (site 11) and one in system 1 (site 9), also tells us that the amplitude at the edge of the system (i.e., $|\Psi_9|$) is equal to $t/t'$ times the constant $|\Psi_n|$ in the reservoir. In the case shown in Fig.~\ref{fig:2systems}(d), the constant-intensity mode is achieved at $t'\approx1.01t$, and indeed we find $|\Psi_{10}|/|\Psi_9|$ is given by the same ratio 1.01. This property indicates that in the weak coupling regime where $0<t'\ll t$, $I_c$ in the reservoir drops significantly as $|t'/t|^2$ compared to the peak intensity in the two systems.

\section{Other Lucas sequences}

\begin{figure}[t]
  \centering \includegraphics[width = \linewidth]{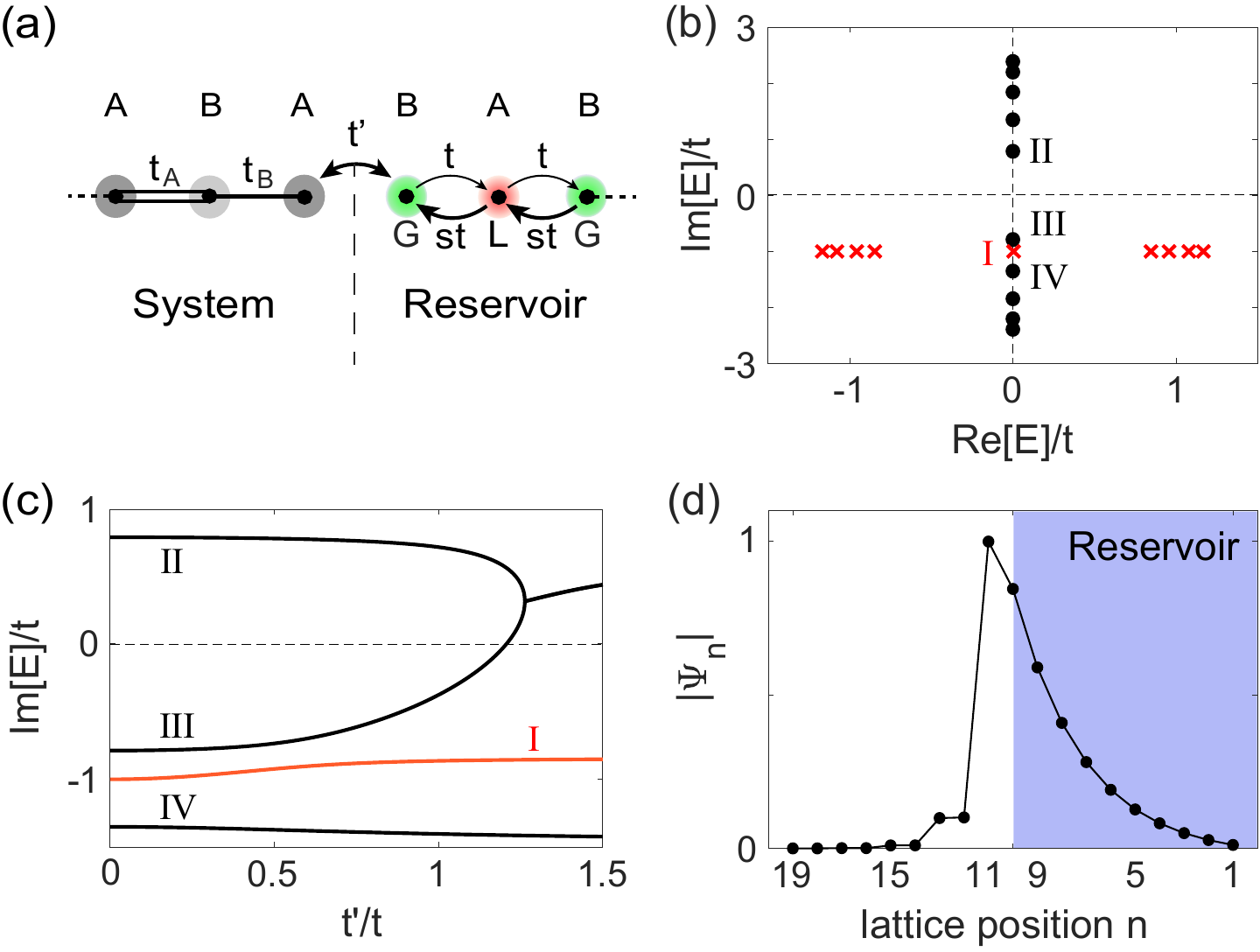}
  \caption{Same as Fig.~\ref{fig:schematic} but with asymmetric couplings $t$ and $st$ $(s>0)$ in the non-Hermitian reservoir. Here $P=3$, $Q=2$, $s=\sqrt{2}$, and $\gamma = 1+\sqrt{2}$. }\label{fig:generalize}
\end{figure}

Our discussion so far has focused on Eq.~(\ref{eq:iter2}), where $P=2$ and $Q=1$. If we introduce asymmetric couplings \cite{Hatano} in the non-Hermitian reservoir on top of the alternate gain and loss \cite{NHchiral} [see Fig.~\ref{fig:generalize}(a)], Eq.~(\ref{eq:iter2}) now takes the general form (\ref{eq:lucas}) that defines the Lucas sequences, with
\be
P = 2\alpha = \frac{\gamma^2}{t^2} - 2s,\quad Q = s^2.\label{eq:gen}
\ee
$s$ is the asymmetric coupling ratio in the reservoir, and below we first take it to be real. This restriction maintains the NHPH symmetry of the coupled lattice, from which we can easily obtain a zero mode with $E=0$ as outlined before.

For the convenience of discussing the open boundary condition on the right side of the reservoir, we have relabeled the lattice sites in Eq.~(\ref{eq:gen}), i.e., we count from the rightmost one in the reservoir and label it site 1 [see Fig.~\ref{fig:generalize}(d)]. It is apparent that this relabeling has no effect on the recurrence relation (\ref{eq:lucas}) in the $s=1$ case studied in previous sections, because this relation is unchanged after switching the two terms $\Psi_n$ and $\Psi_{n-4}$. In the $s\neq1$ case though, $P$ and $Q$ would take different forms from those given by Eq.~(\ref{eq:gen}) without this relabeling. 

As mentioned previously, the open boundary condition here is equivalent to the Dirichlet boundary condition on a corresponding lattice with an added site on the right (i.e., $\Psi_0=0$ at site 0) \cite{Ge_PTlength}. Therefore, the sublattice with even-numbered sites must give the Lucas sequence $U_m$ in Eq.~(\ref{eq:Vm}), with $\Psi_0$ pinned at 0 by the effective Dirichlet boundary condition and $\Psi_2$ scaled to 1: 
\be 
\Psi_{2m} = U_m\quad(m=0,1,2,\ldots) \label{eq:sublatticeA}
\ee
For the odd-numbered sites, the open boundary condition on the right side of the non-Hermitian reservoir leads to 
\be
(P+s) \Psi_1 = \Psi_3, \label{eq:sublatticeB}
\ee
which means the sequence on this sublattice is a combination of the two complementary Lucas sequences: 
\be
\Psi_{2m+1} =  (P+2s)U_m + V_m \quad(m=0,1,2,\ldots) \label{eq:mix}
\ee
One may wonder why this simultaneous excitation does \textit{not} take place in the $P=2$ and $ Q=1$ case too, which would destroy linear localization. The reason is that Eq.~(\ref{eq:mix}) assumes that the Lucas sequence(s) exist on each sublattice. The $P=2$ and $ Q=1$ case, however, is special in that the Lucas sequences are realized by treating the entire reservoir as its sublattice. For example, Eq.~(\ref{eq:sublatticeB}) now become $\psi_3=3\psi_1$ with $P=2$ and $ Q=1$. Instead of treating $\Psi_1,\Psi_3$ as two consecutive numbers in a sequence, we include $\Psi_2$ as their intermediate. Therefore, we can write $|\Psi_n| = n\,(n=1,2,\ldots)$ that satisfies both Eq.~(\ref{eq:sublatticeB}) and (\ref{eq:sublatticeA}), with the latter scaled by a factor of 2 (i.e., $\Psi_{2m}=2m$).     

Despite this mixing of the two complementary Lucas sequences $U_m$ and $V_m$ on one sublattice in general, the latter can still be reconstructed using 
\be
V_m =  \frac{2\Psi_{2m+1}}{\Psi_1} - (P+2s)\frac{\Psi_{2m+1}}{\Psi_2}.  
\ee
As an example, let us consider the case with $P=3$ and $Q=2$. The corresponding Lucas sequence 
\be 
U_m = 2^m-1 
\ee 
is known as the Mersenne numbers, while the complementary Lucas sequence 
\be 
V_m = 2^m+1 
\ee 
includes the Fermat numbers $2^{2^m}+1$. Using Eq.~(\ref{eq:gen}), we set the asymmetric coupling ratio $s=\sqrt{2}$ and the gain and loss modulation $\gamma=1+\sqrt{2}$ in the non-Hermitian reservoir. 

Similar to the $s=1$ case shown in Fig.~\ref{fig:schematic}, we let the uniform loss in the system on the left be $\kappa_0=t$ [see the red crosses in Fig.~\ref{fig:generalize}(b)], and a zero mode with $E=0$ is obtained at $t'=1.207t$ [see Fig.~\ref{fig:generalize}(c)]. We indeed confirm that in the reservoir
\be
\frac{\psi_\text{even}}{\psi_2}=0,1,3,7,15,31\nonumber
\ee
(including $\Psi_0=0$) are given by the Mersenne numbers $U_m$, and 
\be 
\frac{2\psi_\text{odd}}{\psi_1} - (P+2s)\frac{\psi_\text{even}}{\psi_2} = 2,  3,  5,   9, 17  \nonumber
\ee  
gives its complementary Lucas sequence $V_m$.  

\begin{figure}[b]
  \centering \includegraphics[width = \linewidth]{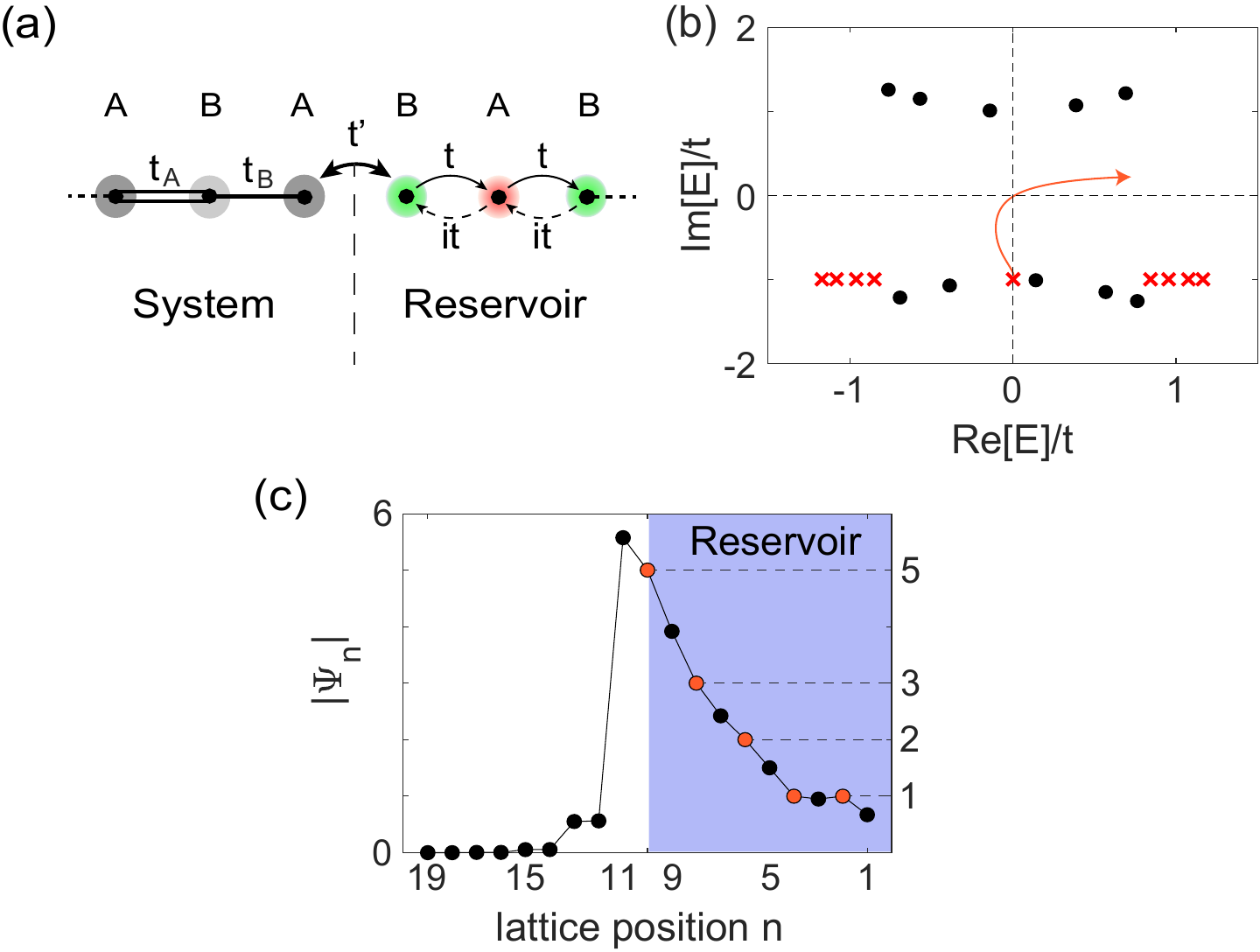}
  \caption{(a) Same as Fig.~\ref{fig:generalize}(a) but with a negative $Q=-1$ (and an imaginary $s=i$) and $P=1$ in the non-Hermitian reservoir. $\gamma=\sqrt{1+2i}$ indicates both gain and loss modulation and on-site detuning. (b) Spectra of the system (red crosses) and the reservoir (black dots) when $t'=0$. Red line with arrow shows the motion of the original Hermitian zero mode as $t'$ increases to $1.5t$. It almost crosses the origin at $t'\approx1.137t$. (c) Wave function of this near zero mode. Shaded area indicates the reservoir. The Fibonacci numbers on one sublattice are highlighted by colored dots and marked on the right vertical axis. }\label{fig:Fibonacci}
\end{figure}

As we mentioned at the beginning of this section, we have required a real $s$ in Eq.~(\ref{eq:gen}) such that NHPH symmetry still holds, and in turn, the spectrum of the coupled system-reservoir is symmetric above the imaginary axis. This approach provides a single parameter tuning (i.e., $t'$) to realize a zero mode with $E=0$. If we lift this restriction and allow $s$ to be imaginary (and $Q$ negative), we can in principle realize \textit{all} Lucas sequences on our platform, including the Fibonacci numbers (with $P=1=-Q$). This most general case, however, requires at least two-parameter tuning to achieve a zero mode (e.g., the uniform loss $\kappa_0$ in the system besides the coupling $t'$). If such a zero mode can be achieved, we then find that Eqs.~(\ref{eq:sublatticeA}) and (\ref{eq:mix}) still hold, and we can again obtain $U_m$ on one sublattice and reconstruct $V_m$ using wave function on both sublattices. 

As an example, consider the $P=1=-Q$ case that leads to the Fibonacci and Lucas numbers. Using Eq.~(\ref{eq:gen}), we set the asymmetric coupling ratio $s=i$ [see Fig.~\ref{fig:Fibonacci}(a)], and the complex modulation $\gamma=\sqrt{1+2i}$ means that there is an alternate on-site detuning of $\pm\re{\gamma}$ on top of the gain and loss modulation. 

We again let the uniform loss in the system on the left be $\kappa_0=t$, and an approximate zero mode with $E=0.012t$ is obtained at $t'=1.137t$ [see Fig.~\ref{fig:Fibonacci}(b)]. We indeed confirm that in the reservoir
\be
\frac{\psi_\text{even}}{\psi_2}\approx0,1.000,1.000,2.000,3.000,5.002\nonumber
\ee
(including $\Psi_0=0$) closely represent the Fibonacci numbers $U_m$ [see Fig.~\ref{fig:Fibonacci}(c)], and 
\be 
\frac{2\psi_\text{odd}}{\psi_1} - (P+2s)\frac{\psi_\text{even}}{\psi_2} \approx 2.000, 1.000, 3.000, 4.001, 7.002\nonumber
\ee  
gives the complementary Lucas numbers $V_m$. The minimum of $|E|$ can be further reduced by one order of magnitude, for example, if we let $\kappa_0=0.2t$ and $t'=0.512t$ (i.e., two-parameter tuning), with which the approximate Fibonacci and Lucas numbers given above differ from integers by less than $2\times10^{-5}$.

\section{Conclusion}

In summary, we showed that the two complementary Lucas sequences of a broad range can be observed on the same platform, using zero modes in coupled photonic lattices or similar systems. Such a photonic lattice can be realized, for example, using coupled on-chip micro-ring resonators \cite{Gao2023,Feng2025} with a typical size of tens of microns and photonic crystal defect cavities \cite{Ji2023,Ji2025} smaller by one order of magnitude.   
III-V multi-layer quantum wells are commonly used as the gain medium, which can also be applied to on-chip fabricated ridge waveguides with a cross section several hundred nanometers along each dimension \cite{Guo} (see realistic parameters given in Ref.~\cite{LL}). Acoustic waveguides \cite{Chen2022} and cavities \cite{Ding2016} can also be utilized to form the required lattice, and they are easier to fabricate given their tabletop sizes, where couplings are realized by slots or tubes in the same metallic block. Last but not least, electrical circuits have also emerged as an easily configurable platform to realize tight-binding Hamiltonians, which can be fit onto a single breadboard \cite{Imhof}. 

In the special case of $P=2$ and $Q=1$, these complementary Lucas sequences manifest themselves as linear localization and a constant-intensity mode respectively, 
in a gain and loss modulated reservoir bridging two mirror-symmetric systems. More generally, one Lucas sequence ($U_m$) can be observed directly on a sublattice of the reservoir, when the latter is terminated with the open boundary condition on one side. Its complementary Lucas sequence $V_m$, on the other hand, can be reconstructed using the wave function on both sublattices in the reservoir. 

To fulfill the conditions for the recurrence relation that defines the Lucas sequences with $Q>0$, we resorted to NHPH symmetry that warrants non-Hermitian zero modes on the imaginary energy axis, followed by properly choosing the coupling between the reservoir and the two systems on its sides. By including loss in the two systems, this coupling can be made comparable or even stronger than the coupling in the reservoir, leading to pronounced linear tail and constant intensity in the reservoir. 

Unlike previously identified constant-intensity modes, our instance requires only modulating the imaginary part of the on-site potential. In addition, the phase of this mode is also a constant on each sublattice of the reservoir, leading to an equal division of energy flux from a gain site to its two neighboring loss sites. Through the discussion of different configurations, we also demonstrated that linear localization can persist with two different systems bridged by the reservoir and be manifested by multiple zero modes. We also note that while our constant-intensity mode may look similar to a flat-top beam \cite{flattop}, the latter is created using refractive or reflective optical elements and not a resonant mode. 

To realize Lucas sequences with $Q<0$, we required alternate on-site frequency detunings besides the gain and loss modulation and asymmetric couplings in the reservoir. While NHPH symmetry no longer holds, a multi-parameter scanning can in principle produce the two Lucas sequences, as we demonstrated using the Fibonacci and Lucas numbers in the case of $P=1=-Q$.

Although we mainly focused on a reservoir with both gain and loss, these Lucas sequences, including linear localization and constant-intensity modes, can also be observed in an entirely passive setup with only loss. It only requires imposing additional uniform loss (across the entire lattice) stronger than the gain in the original reservoir. Such a configuration is most valuable in systems where gain is inaccessible or difficult to implement, including silicon waveguides \cite{Pan2018} and laser-written waveguides in fused silica glass \cite{Szameit}.

Our discussion also sheds light on achieving different effective boundary conditions at the interface of two coupled lattices, which may help achieve other exotic states in coupled systems with potential topological protection.

\begin{acknowledgments}
This project is supported by the National Science Foundation (NSF) under Grant No. DMR-2326698.
\end{acknowledgments}

\end{document}